\documentstyle[psfig,12 pt]{article}
\begin{document}
\textwidth=16 cm
\hsize=6.0 true in \hoffset=.25 true in

\begin{center}
{\Large{\bf Two-site two-electron Holstein model:\\
 a perturbation study }}\\
\end{center}
\vskip 1.0cm
\begin{center}
 Jayita Chatterjee\footnote{ e-mail: moon@cmp.saha.ernet.in} 
 and A. N. Das
\end{center}
\vskip 0.50cm
\begin{center}
 {\em Theoretical Condensed Matter Physics Group \\
 \em Saha Institute of Nuclear Physics \\
1/AF Bidhannagar, Kolkata 700064, India}\\
\end{center}

\vskip 1.0cm

\vskip 1.0cm
\begin{center}
{\bf Abstract}
\end{center}
\vskip 0.5cm
 The two-site two-electron Holstein model is studied within a 
perturbation method based on a variational phonon basis obtained 
through the modified Lang-Firsov (MLF) transformation. 
The ground-state wave function and the energy are found out 
considering  up to the seventh and eighth order of perturbation, 
respectively.
The convergence of the perturbation corrections
to the ground state energy as well as to the correlation 
functions are investigated. 
The kinetic energy and the correlation functions involving charge and 
lattice deformations are studied as a function of electron-phonon 
($e$-ph) coupling and electron-electron interactions for different
values of adiabaticity parameter.
The simultaneous effect of the $e$-ph coupling and Coulomb repulsion 
on the kinetic energy shows interesting features.  

\vskip 1.5cm

{\bf 1. Introduction}
\vskip 0.5cm

The Holstein model \cite {Hol} is one of the fundamental models 
describing the interactions of conduction electrons with lattice 
vibrations and has been studied widely for a long time using 
different analytical and numerical methods. However, conventional analytical 
methods based on variational and perturbation approach are found 
to be much less satisfactory than the numerical methods; particularly 
in a wide region of intermediate values of electron-phonon ($e$-ph) 
coupling and hopping.
The importance of a reliable analytical approach is crucial due to
the failure of standard analytical approximations on the Holstein model;
the Migdal
approximation \cite{Mig} is valid for a weakly
coupled adiabatic system, while the Lang-Firsov canonical transformation
\cite{LF} could not be suitable far from the nonadiabatic
limit and strong coupling limit.
The failure of the standard analytical techniques in the
intermediate range of coupling has been observed by many
workers \cite{Feh} while comparing with exact results.
Recently we \cite{DJC} developed a perturbative expansion 
using different phonon bases obtained through the LF, MLF and MLF with 
squeezing (MLFS) transformations for a two-site single electron 
Holstein model. We found that the perturbation corrections within the 
MLF and MLFS methods are much smaller than the corresponding LF 
values for weak and intermediate couplings. The MLF method shows 
a very satisfactory convergence in perturbation expansion and is 
able to produce almost exact results for the entire region of 
the $e$-ph coupling from weak to intermediate values of hopping.

Our aim is to extend the applicability of our approach,
developed for a single electron system, to a many-electron system.
To examine that whether this perturbation method based on MLF phonon
basis works well or not with increasing number of electrons and in presence
of electron-electron ($e$-$e$) interaction,
we consider a two-site two-electron Holstein model 
which is the minimal model to investigate the effect of
$e$-$e$ interaction on the polaronic properties and the combined 
effects of the electronic correlation and the $e$-ph coupling on the 
kinetic energy of the system. It may be mentioned that the polaron crossover
has been studied in two-site and four-site systems in presence of
$e$-$e$ interaction by different groups \cite{PF,DC,HH} with the Holstein
model previously. But most of the studies were done at the zero phonon
approximation level with MLF or MLFS transformed
Hamiltonian. However, the combined effect of $e$-$e$ interaction and
$e$-ph interaction on the kinetic energy has not been properly studied
and this is the main motivation of the present study.

In our study, we have developed a perturbation series 
using the MLF phonon basis and 
examined the convergence by comparing contributions of different 
orders of perturbation to the energy and the correlation functions
for the ground state.   
The variation of different correlation functions involving charge
and lattice deformation and the kinetic energy of the system with
$e$-ph coupling are presented and discussed for different values of
adiabaticity parameters and Coulomb repulsion.
\vskip 1.0cm

{\bf 2. Model Hamiltonian and formalism}
\vskip 0.5cm

The two-site two-electron Holstein Hamiltonian is
\begin{eqnarray}
H &=& \sum_{i,\sigma} \epsilon n_{i \sigma} - \sum_{\sigma}
t (c_{1 \sigma}^{\dag} c_{2 \sigma} 
+ c_{2 \sigma}^{\dag} c_{1 \sigma})
+U\sum_{i}n_{i\uparrow}n_{i\downarrow} + V~n_1 n_2 
\nonumber \\
&+& g_1 \omega_0  \sum_{i,\sigma}  n_{i \sigma} (b_i + b_i^{\dag}) 
+ g_2 \omega_0  \sum_{i,\sigma}  n_{i \sigma} (b_{i+\delta} 
+ b_{i+
\delta}^{\dag}) 
+  \omega_0 \sum_{i}  b_i^{\dag} b_i 
\end{eqnarray} 
where $i$ =1 or 2, denotes the sites. $c_{i\sigma}$ ($c_{i\sigma}^{\dag}$)  
is the annihilation (creation) operator for the electron with spin 
$\sigma$ at site $i$ and $n_{i \sigma}$ (=$c_{i\sigma}^{\dag} c_{i\sigma}$) 
is the corresponding number operator, 
$g_1$ and $g_2$ denote, respectively, the on-site and inter-site
$e$-ph coupling  strengths. 
$t$ is the usual hopping integral. $b_i$ ($b_{i}^{\dag}$) is the annihilation
(creation) operator for the phonons corresponding to 
interatomic vibrations at site $i$, $\omega_0$ is the phonon frequency, 
$U$ and $V$ are, respectively, the on-site and inter-site Coulomb 
repulsion between the electrons. 

Introducing new phonon operators ${\bf a}=~(b_1+b_2)/ \sqrt 2$ and 
$d=~(b_1-b_2)/\sqrt 2 $ the Hamiltonian is separated into two parts   
($H=H_d + H_{{\bf a}})$ \cite{DJC}: 
\begin{eqnarray}
H_d &=& \sum_{i,\sigma} \epsilon n_{i, \sigma} - 
\sum_{\sigma} t (c_{1 \sigma}^{\dag} c_{2 \sigma} 
+ c_{2 \sigma}^{\dag} c_{1 \sigma}) 
+U\sum_{i}n_{i\uparrow}n_{i\downarrow} + V~n_1 n_2 \nonumber \\
&+& \omega_0  g_{-} (n_1-n_2) (d + d^{\dag}) 
+  \omega_0  d^{\dag} d 
\end{eqnarray}
and  
\begin{equation}
H_{\bf a} =  \omega_0 \tilde{{\bf a}}^{\dag}\tilde{{\bf a}} - \omega_0 n^2 g_{+}^2 
\end{equation}
where $g_{+}=(g_1 + g_2)/\sqrt 2$, $g_{-}=(g_1 - g_2)/\sqrt{2}$, 
$\tilde{{\bf a}}={\bf a} + ng_{+}$, $\tilde{{\bf a}}^{\dag}={\bf a}^{\dag} +ng_{+}$, 
$n=n_1+n_2$ and $n_i=n_{i\uparrow}+n_{i\downarrow}$.

$H_{\bf a} $ describes a shifted oscillator which couples only  
with the total number of electrons $n$, which 
is a constant of motion. $H_d$ represents an effective $e$-ph
system where phonons directly 
couple with the electronic degrees of freedom. 
We now use the MLF transformation where the lattice deformations 
produced by the electron are treated as variational parameters 
\cite {DC,DS,LS}. For the present system the relevant transformation 
is, 
\begin{equation}
\tilde{H_d} = e^R H_d e^{-R}
\end{equation}
where $R =\lambda (n_1-n_2) ( d^{\dag}-d)$, $\lambda$ is a
variational parameter related to the displacement of the $d$ 
oscillator. 

The transformed Hamiltonian is then obtained as \cite{DC} 
\begin{eqnarray}
\tilde{H_d}&=&\omega_0  d^{\dag} d + \sum_{i,\sigma} \epsilon_p n_{i \sigma} 
- \sum_{\sigma} t [c_{1 \sigma}^{\dag} c_{2 \sigma}~ 
\rm{exp}(2 \lambda (d^{\dag}-d))    
+ c_{2 \sigma}^{\dag} c_{1 \sigma}~\rm{exp}(-2 \lambda (d^{\dag}-d))]
\nonumber  \\
&+& \omega_0  (g_{-} -\lambda) (n_1-n_2) (d + d^{\dag}) 
+U_e\sum_{i}n_{i\uparrow}n_{i\downarrow} + V_e~n_1 n_2  
\end{eqnarray}
where 
\begin{eqnarray}
\epsilon_p &=& \epsilon - \omega_0 ( 2 g_{-} - \lambda)~\lambda \\
U_e &=& U - 2\omega_0 ( 2 g_{-} - \lambda)~\lambda \\
V_e &=& V + 2\omega_0 ( 2 g_{-} - \lambda)~\lambda 
\end{eqnarray}

For the two-site two-electron system there are six electronic 
states of which three belong to the triplet states as follows:
\begin{eqnarray}
|\psi_{T1} \rangle &=&   c_{1 \uparrow}^{\dag}   
c_{2 \uparrow}^{\dag} |0\rangle_{e}   \\
|\psi_{T2} \rangle &=&   \frac{1}{\sqrt 2}~( 
c_{1 \uparrow}^{\dag}    c_{2 \downarrow}^{\dag}    
+c_{1 \downarrow}^{\dag}   c_{2 \uparrow}^{\dag} )~|0\rangle_{e}\\ 
|\psi_{T3} \rangle &=&   c_{1 \downarrow}^{\dag}   
c_{2 \downarrow}^{\dag} |0\rangle_{e}   
\end{eqnarray}
and other three states are singlet as given below: 
\begin{eqnarray}
|\psi_{1} \rangle &=&   \frac{1}{\sqrt 2}~( 
c_{1 \uparrow}^{\dag}   c_{1 \downarrow}^{\dag}  
-  c_{2 \uparrow}^{\dag}  c_{2 \downarrow}^{\dag}
)~|0\rangle_{e} \\ 
|\psi_{2} \rangle &=&   \frac{1}{\sqrt 2} ~(
c_{1 \uparrow}^{\dag}    c_{1 \downarrow}^{\dag}  
+  c_{2 \uparrow}^{\dag}  c_{2 \downarrow}^{\dag}
)~|0\rangle_{e} \\ 
|\psi_{3} \rangle &=&   \frac{1}{\sqrt 2}~( 
c_{1 \uparrow}^{\dag}   c_{2 \downarrow}^{\dag}     
- c_{1 \downarrow}^{\dag}    c_{2 \uparrow}^{\dag})~ |0\rangle_{e}  
\end{eqnarray}

The triplet states do not mix with each other or couple with the 
$d$-oscillator. In the following we will only consider the singlet 
states and the $d$-oscillator which are mutually coupled through 
the off-diagonal terms of the Hamiltonian (5).    
 For a perturbation method it is desirable to use a basis 
where the major part of the Hamiltonian becomes diagonal. 
For this purpose we choose the basis set $|\psi_{+},N \rangle$,  
$|\psi_{1},N \rangle$ and $|\psi_{-},N \rangle$ 
defined as
\begin{eqnarray}
|\psi_{+},N \rangle &=& \frac{1}{\sqrt {a_{N}^2+b_{N}^2}} 
(a_{N} |\psi_2 \rangle  +b_{N} |\psi_3\rangle )  |N\rangle \\
|\psi_{1},N \rangle &=& |\psi_{1}\rangle |N \rangle \\  
|\psi_{-},N \rangle &=& \frac{1}{\sqrt {a_{N}^2+b_{N}^2}} 
(a_{N} |\psi_3\rangle  -b_{N} |\psi_2 \rangle )  |N\rangle 
\end{eqnarray}
where $|\psi_+\rangle$, $|\psi_1\rangle$ and $|\psi_-\rangle$ 
represent the bonding, 
nonbonding and antibonding electronic states, respectively and 
$|N\rangle$ denotes 
the $N$th excited oscillator state in the MLF phonon basis where 
$N$ takes integer values from 0 to $\infty$.\\
Here $a_N= - (U_e - V_e) +\sqrt{ (U_e - V_e)^2 +16 t^2_e(N,N) }$
and $b_N= {4 t_e(N,N)} $.\\
where $t_{e}(N,N) = t_e \sum_{p=1}^{N} (-1)^p \frac{(2 \lambda)^{2p}}
{p!}~ N_{C_p} $,  $N_{C_p}=\frac{N!}{p!(N-p)!}$,\\
and $t_e=t~\rm{exp}{(-2\lambda^2)}$.  

The hopping term in Eq. (5)
has both diagonal and offdiagonal
matrix elements in the chosen basis (Eqs. 15-17). 
The states $|\psi_{e},N \rangle$ (for $e=+,1,-$) are
the eigenstates of the unperturbed Hamiltonian   
\begin{eqnarray}
\tilde{H_0}&=&\omega_0 d^{\dag} d + \sum_{i} \epsilon_p n_{i} - 
\sum_{\sigma}t_e ~ F(d^{\dag}d)~ [c_{1 \sigma}^{\dag} c_{2 \sigma} 
+ c_{2 \sigma}^{\dag} c_{1 \sigma}] \nonumber  \\
&+& U_e\sum_{i}n_{i\uparrow}n_{i\downarrow} + V_e~n_1 n_2  
\end{eqnarray}
It may be noted that $t_{e}F(d^{\dag}d)$ in the above Eq. (18) is the
diagonal part of the operator $t~\exp [\pm 2\lambda (d^{\dag}_i -d_i)]$
in Eq. (5) in the phonon occupation number representation. 
\begin{eqnarray}
{\rm Since}~  \exp [2\lambda (d^{\dag} -d)] &=& 
~  \exp[2\lambda d^{\dag}] ~\exp[ -2\lambda d] \exp[
\frac{4\lambda^2}{2} [d^{\dag} ,d]_{-} ] \nonumber \\
&=& e^{-2\lambda^2 }   ~  \sum_p \sum_q (-1)^q ~\frac{(2\lambda)^{p+q} }{p!~q!}
  ~(d^{\dag})^{p} (d)^{q}  , \nonumber\\
{\rm so}~ F(d^{\dag}d) &=& ~\sum_{p=1}^N (-1)^p~ (2 \lambda)^{2p}
(d^{\dag}d)^{p}/ (p!)^2.\nonumber  
\end{eqnarray}
For the choice of the unperturbed Hamiltonian in Eq. (18)
the perturbation Hamiltonian ($\tilde{H} -\tilde{H_{0}}$) has no
diagonal matrix element in our chosen basis. This is not achieved when
one replaces $ F(d^{\dag}d)$ by 1 and $a_N$, $b_N$ become $N$-independent.
The convergence in perturbation expansion for the latter case
is found to be weaker than that when $ F(d^{\dag}d)$ is included
in $H_0$.

The unperturbed energies corresponding to the eigenstates
$|\psi_{e},N \rangle$ ($e=+,1,-$) are 
\begin{eqnarray}
E_{+,N}^{(0)} 
&=& N \omega_0  + 2 \epsilon_p +\frac{(U_e+V_e)}{2} - \frac{1}{2}
\sqrt{(U_e-V_e)^2 + 16 t_{e}^{2}(N,N) },\nonumber \\
E_{1,N}^{0}&=& 2 \epsilon_p +U_e \nonumber\\
 \rm{and}~  E_{-,N}^{(0)} 
&=& N \omega_0  + 2 \epsilon_p +\frac{(U_e+V_e)}{2} + \frac{1}{2}
\sqrt{(U_e-V_e)^2 + 16 t_{e}^{2}(N,N) } 
\end{eqnarray}

The part of the Hamiltonian, which is not included in $\tilde{H_0}$, is 
taken as the perturbation $\tilde{H_1}$.   
For the chosen basis, the state $|\psi_+,0\rangle$ has the lowest
unperturbed energy $E_{+,0}^{(0)}$. The general off-diagonal matrix 
elements of 
$\tilde{H_1}$ between the states $|\psi_{\pm},N \rangle$ and $|\psi_{1},M 
\rangle$ 
are calculated as (for $(N-M)>0$)
\begin{eqnarray}  
\langle N,\psi_{\pm}|\tilde{H_{1}}|\psi_{\pm}, M \rangle &=& \mp
2t_{e}(N,M)~ (a_{N}^{'}b_{M}^{'} + a_{M}^{'}b_{N}^{'})
~~~~~~\rm{for}~\rm{even}~(N-M).\nonumber \\ 
&=& 0 ~~~~~~~~~~~~~~~~~~~~\rm{for}~\rm{odd}~(N-M). \\ 
\langle N,\psi_{\pm}|\tilde{H_{1}}|\psi_{\mp},M \rangle &=& -  
2t_{e}(N,M)~ (a_{N}^{'}a_{M}^{'} - b_{M}^{'}b_{N}^{'})
~~~\rm{for}~\rm{even}~(N-M).\nonumber \\ 
&=& 0 ~~~~~~~~~~~~~~~~~~~~\rm{for}~\rm{odd}~(N-M). \\ 
\langle N,\psi_{1}|\tilde{H_{1}}|\psi_{+},M \rangle &=& 2a_{M}^{'}
 P(N,M)-  2t_{e}(N,M)~ b_{M}^{'}
~\rm{for}~\rm{odd}~(N-M).  \nonumber \\ 
&=& 0 
~~~\rm{for}~\rm{even}~(N-M). \\ 
\langle N,\psi_{1}|\tilde{H_{1}}|\psi_{-},M \rangle &=& -2b_{M}^{'}
P(N,M)-  2t_{e}(N,M)~ a_{M}^{'}
~\rm{for}~\rm{odd}~(N-M).\nonumber \\ 
&=& 0  
~~~~~\rm{for}~\rm{even}~(N-M). \\ 
{\rm where}~~~ t(N,M)&=& t_{e} (2\lambda)^{N-M} 
\sqrt{\frac{N!}{M!}} \left[  \frac{1}{(N-M)!}+\sum_{R=1}^{M}
[(-1)^R \right. \nonumber \\
&& \left. \frac{(2\lambda)^
{2R}}{(N-M+R)! R!}M(M-1)...(M-R+1) ] \right] \nonumber\\
 P(N,M) &=& \omega_0 (g_{-}-\lambda) [\sqrt{M}
~\delta_{N,M-1} + \sqrt{M+1} ~\delta_{N,M+1}] \nonumber  \\
a_{N}^{'} &=&\frac{a_{N}}{\sqrt{(a_{N}^2+b_{N}^2)}} ~{\rm and}~
b_{N}^{'} =\frac{b_{N}}{\sqrt{(a_{N}^2+b_{N}^2)}} \nonumber
\end{eqnarray}  

Now one has to make a proper choice of $\lambda$ so 
that the perturbative expansion is satisfactorily convergent. 
In our previous work we obtained good convergence of the 
perturbation series for the value of $\lambda$ 
which minimizes the unperturbed ground state energy. Following 
the same procedure, i.e. minimizing $E_{+,0}^{(0)}$ 
with respect to $\lambda$ we obtain 
\begin{eqnarray}
\lambda=g_{-}/[1+\frac{4t_{e}}{\omega_0[\sqrt{r^2+4} -r]}  ]
\end{eqnarray}
where $r=(U_e-V_e)/2 t_e$.

The corrected ground state wave function may be written as,
\begin{eqnarray}  
|\psi_{G} \rangle  &=& \frac{1}{\sqrt{N_G}}\left[|\psi_+,0\rangle 
 +\sum_{N=1,3,..}C_{N}^{1} |\psi_1,N \rangle
 +\sum_{N=2,4,..}C_{N}^{\pm}|\psi_{\pm},N \rangle  \right]\\
{\rm where}~~ N_G &=& 1+ \sum_{N=1,3,..}(C_N^1)^2 +\sum_{N=2,4,...}
(~(C_N^+)^2
+(C_N^-)^2~) ~{\rm~ is~ the~ normalization~ }\nonumber\\
{\rm factor.} \nonumber
\end{eqnarray} 

 The coefficients $C_N^1$, $C_N^+$ and $C_N^-$ for different values of 
$N$ are determined using 
the matrix elements of $\tilde{H_1}$ connecting different unperturbed states 
and their energies following Ref. \cite{DJC}. These Coefficients
are determined up to the seventh order of perturbation while
the ground-state energy 
is found out up to the eighth order of perturbation. 
    
We have also studied the static correlation functions involving  
charge and lattice deformations 
$\langle n_1 u_{1}\rangle_{0}$ and 
$\langle n_1 u_{2}\rangle_{0}$,
where $u_1$ and $u_2$ represents the lattice deformations at sites 1 and 2  
respectively. These correlation functions may be written in terms of 
relevant operators as 
\begin{equation}  
n_{1} u_{1,2} = n_1[-2g_{+} + \frac{({\bf a} + {\bf a}^{\dag})}{2}]
\pm~n_1[-\lambda(n_1-n_2)+ \frac{(d+d^{\dag})}{2}]
\end{equation}
The final form of the correlation functions are obtained as
\begin{eqnarray}
\langle n_{1} u_{1,2} \rangle_{0}&=&  
-2 g_{+} \mp \frac{ 2 \lambda}{N_G} \left[ \frac{a_0^2}{a_0^2+b_0^2} + 
\sum_{N=1,3,5,..} (C_N^{1})^2   \right. \nonumber\\
&+& \left. \sum_{N=2,4,..} \frac{1}{a_N^2+b_N^2} (a_N C_N^{+} - b_N C_N^{-})
^2 ~\right]
\pm\frac{1}{N_G}
\left[\frac{a_0}{\sqrt{a_0^2+b_0^2}}  C_1^{1} 
 \right. \nonumber \\
&+&\sum_{N=2,4,..} \frac{ \sqrt{N} C_{N-1}^1}{ \sqrt{a_N^2+b_N^2}}
 (a_N C_N^{+} - b_N C_N^{-})
 \nonumber \\
&+& \left. \sum_{N=3,5,..} \frac{ \sqrt{N} C_{N}^1}{ \sqrt{a_N^2+b_N^2}}
 (a_{N-1} C_{N-1}^{+} - b_{N-1} C_{N-1}^{-})
 \right]  
\end{eqnarray} 
 The kinetic energy of the system in the ground state $E_{ke}= 
-~t~<\psi_G|$
$ \sum_{\sigma}[c_{1 \sigma}^{\dag} c_{2 \sigma }~$\\ 
$\exp(2 \lambda (d^{\dag}-d))$ 
$+ c_{2 \sigma}^{\dag} c_{1 \sigma}~\exp(-2 \lambda (d^{\dag}-d))]$
$|\psi_G>$, has also been calculated using the ground state
wave function $|\psi_G\rangle$, evaluated up to the seventh order 
of perturbation.
\vskip 1.0cm
{\bf 3. Results and discussions}
\vskip 0.5cm

From the relevant analytical expressions, the quantities of 
our interest have been calculated within the LF and the MLF 
perturbation methods considering 35 phonon states (which is more 
than sufficient for $g_-\leq ~2.5$ in the transformed phonon basis).   
For all numerical calculations we take $\epsilon=0$,
$g_2=0$ (i.e., $g_+$ = $g_-$) in this work. 

\begin{center}
\begin{figure}
\psfig{file=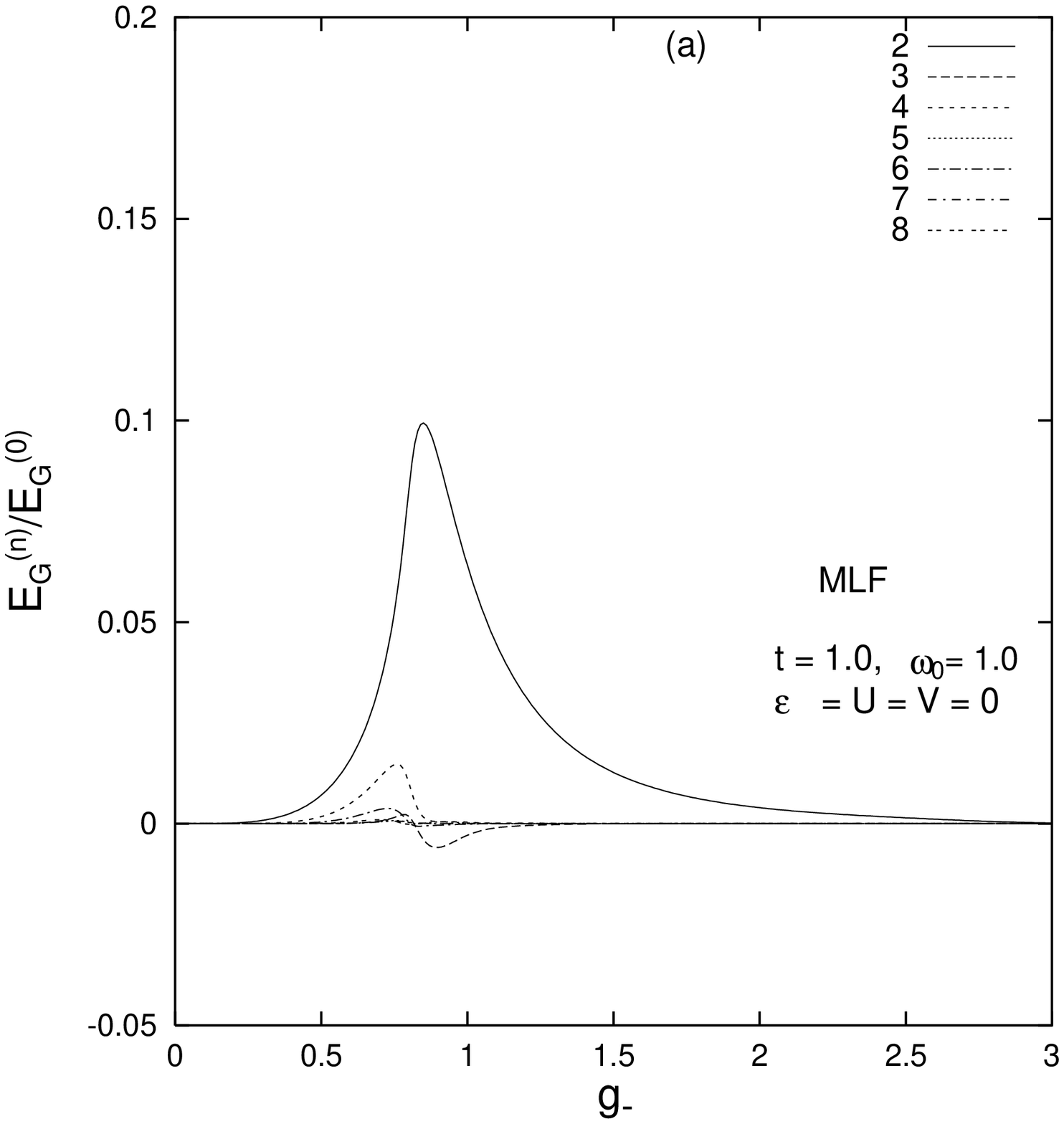,width=3.0in}
\psfig{file=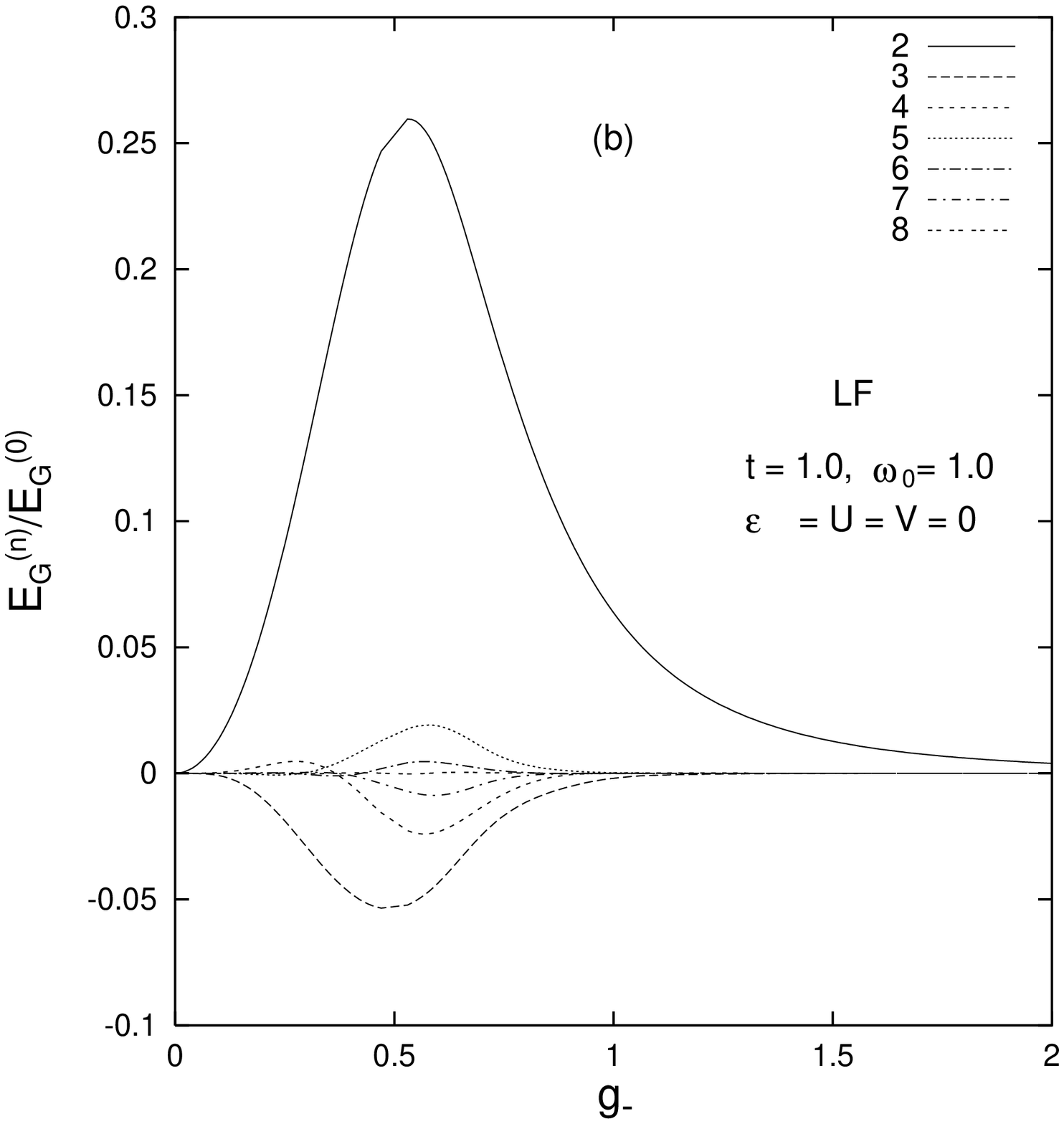,width=3.0in}
\vskip 0.5 cm
\caption{
\label{figure1}
Variation of the relative perturbation corrections 
$E_{G}^{(n)}/E_{G}^{(0)}$ to the ground state energy as a 
function of the coupling strength ($g_{-}$) for $t/\omega_0 =1.0$ 
for (a) MLF method and (b) LF method. 
$E_{G}^{(n)}$ is the nth order perturbation correction to the 
ground state energy and 
$E_{G}^{(0)}$= $E^{(0)}_{+,0}$ 
is the unperturbed ground state energy. 2,..,8 denote the
value of $n$ in $E_{G}^{(n)}$.
}
\end{figure}
\end{center}
\begin{center}
\begin{figure}
\psfig{file=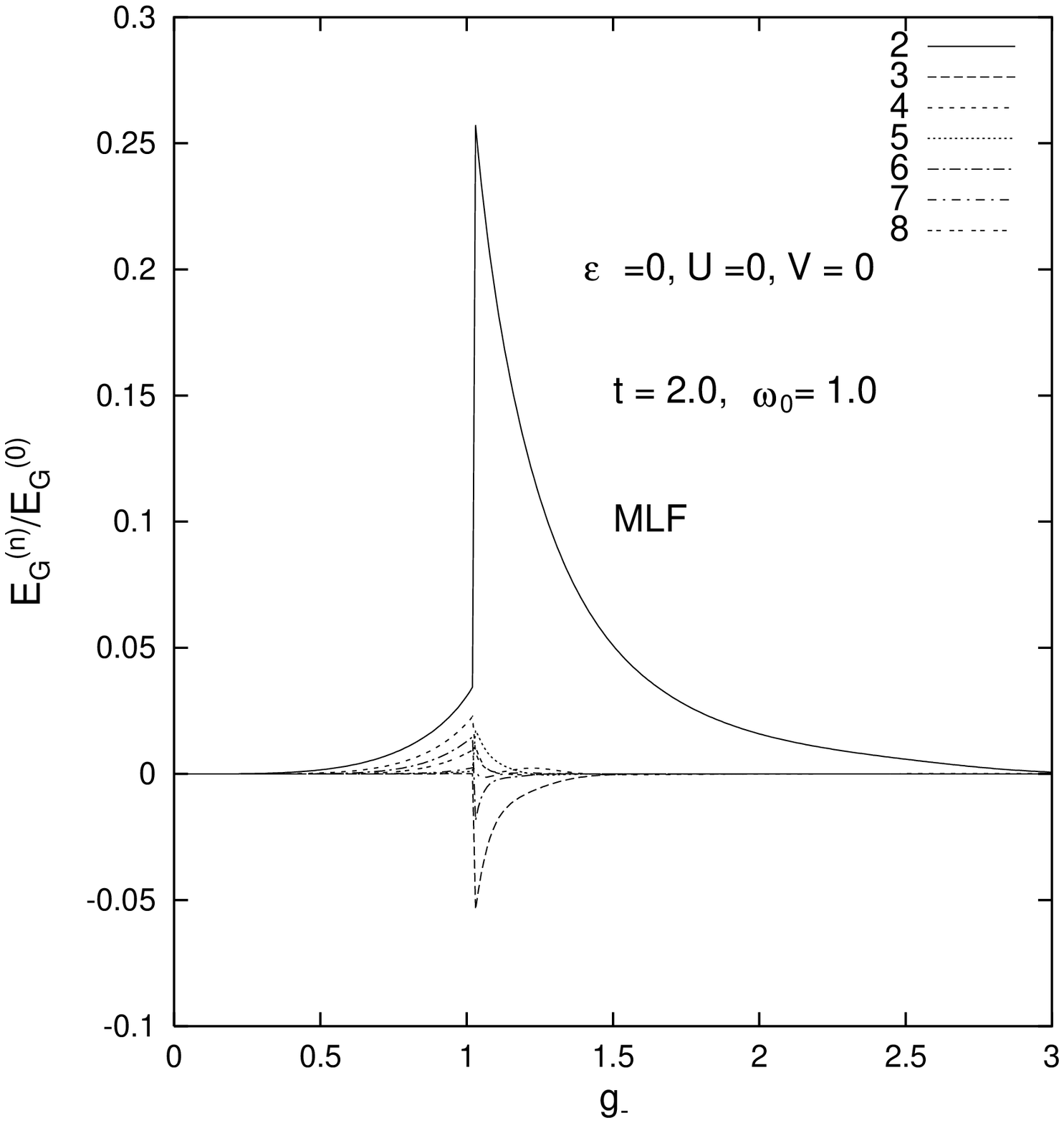,width=3.0in}
\psfig{file=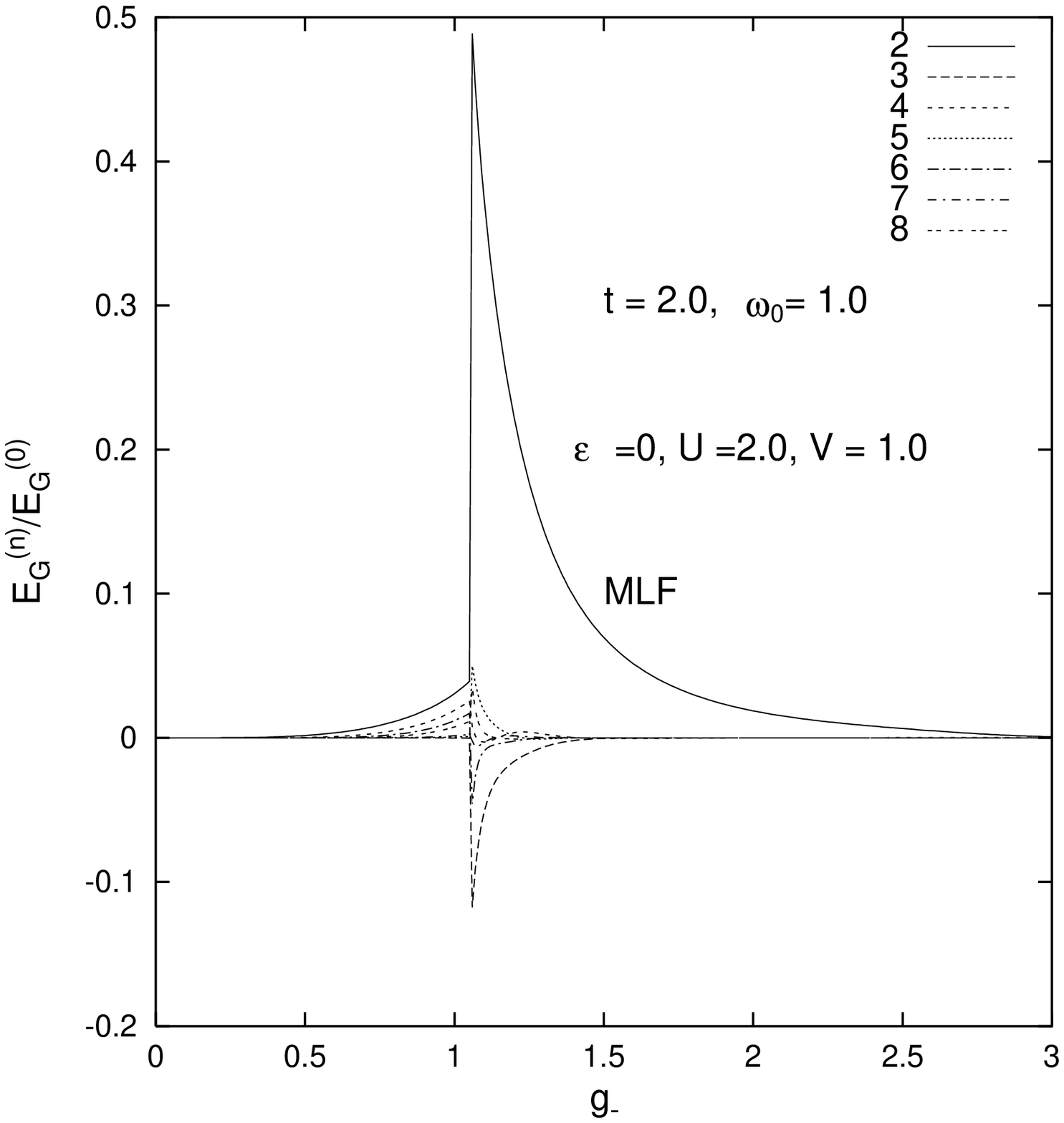,width=3.0in}
\vskip 0.5 cm
\caption{
\label{figure2}
Variation of the relative perturbation corrections 
$E_{G}^{(n)}/E_{G}^{(0)}$ to the ground state energy as a 
function of the coupling strength ($g_{-}$) for $t/\omega_0 =2.0$ 
for the MLF method for (a) $U=0$ and (b) $U=2.0,~V=1.0$. 
$E_{G}^{(n)}$ is the nth order perturbation correction to the 
ground state energy and 
$E_{G}^{(0)}$= $E^{(0)}_{+,0}$ 
is the unperturbed ground state energy. 2,..,8 denote the
value of $n$ in $E_{G}^{(n)}$.
}
\end{figure}
\end{center}
In figure 1 we have shown the relative perturbation corrections to the 
ground-state energy, i.e., the ratios of the perturbation corrections 
of different orders to the unperturbed 
ground-state energy as a function of $g_-$ for $t/\omega_0 =1$. It appears from 
figure 1 that the convergence is more or less satisfactory 
in both the MLF and LF approaches. 
Apart from the second order correction, 
higher order energy corrections are very small within the MLF 
approach except in a small region of $g_- ~\sim~ 0.6$ to 0.8 where 
the convergence is relatively weaker. The energy corrections 
of fifth to eighth order are, however, negligibly small even 
in this region. The energy correction of any order within the LF 
method is larger than the corresponding correction within the MLF 
approach. For smaller values of $t/\omega_0$ the 
corrections are much less and the convergence is better as expected
from our previous study \cite{DJC}.   

 To examine the applicability of our method for higher values of $t$
and also in presence of $e$-$e$ interaction,
we plot in figure 2 the relative perturbation corrections of
different orders to the 
ground-state energy as a function of $g_-$ for
$t/\omega_0 =2$ for both zero and nonzero values of $e$-$e$ interactions.
Figure 2(a) shows that the convergence is satisfactory and
except the second order correction, 
higher order energy corrections are small.
In presence of electronic interactions (nonzero
$U$ and $V$), magnitudes of the relative energy corrections
increase but the convergence is still good (figure 2(b)).
We calculate the ground-state energy within the MLF method  
up to the eighth order of perturbation. As 
pointed out previously, the fifth to eighth order energy corrections are 
so small that the energy calculated up to the eighth order of 
perturbation may be treated as an almost exact result.
We find that our results exactly match with that obtained by the
exact diagonalization study \cite {MR} for the same set of parameters.   

 The on-site and inter-site charge-lattice deformation correlations,
$\langle n_{1} u_{1} \rangle_{0}$  and
$\langle n_{1} u_{2} \rangle_{0}$, are determined for the ground state
considering up to the seventh order of perturbation within the MLF method.
In figure 3 we have plotted these quantities against $g_-$.
When we examine the convergence by comparing the values of the same 
correlation function, calculated up to different orders of perturbation 
we find that the values obtained considering up to the fourth to the seventh 
order of perturbation, remain almost same. This ensures a good convergence 
of our perturbation method for the correlation function and thus 
the results obtained are expected to match satisfactorily
with the exact result.  
The unperturbed LF and the MLF results are also plotted in figure 3 
for a comparison, which shows that the unperturbed MLF results
for the correlation functions are 
much closer to the exact result than the corresponding LF result. 
The relative difference between the exact MLF result and the 
MLF (ZPA) result is much less for 
$\langle n_{1} u_{1} \rangle_{0}$ 
than for $\langle n_{1} u_{2} \rangle_{0}$. This is owing to the fact 
that $\langle n_{1} u_{2} \rangle_{0}$ is a very sensitive function of 
the perturbation corrections to the wave function compared to   
$\langle n_{1} u_{1} \rangle_{0}$. The MLF perturbation result predicts  
a long tail for $\langle n_{1} u_{2} \rangle_{0}$ 
in the strongly coupled region where both the LF (ZPA) 
and the MLF (ZPA) give zero value for  
$\langle n_{1} u_{2} \rangle_{0}$. 
This result is a signature of the retardation effect even in the 
strong coupling region where even the MLF (ZPA) do not predict 
the retardation effect.
\begin{center}
\begin{figure}
\psfig{file=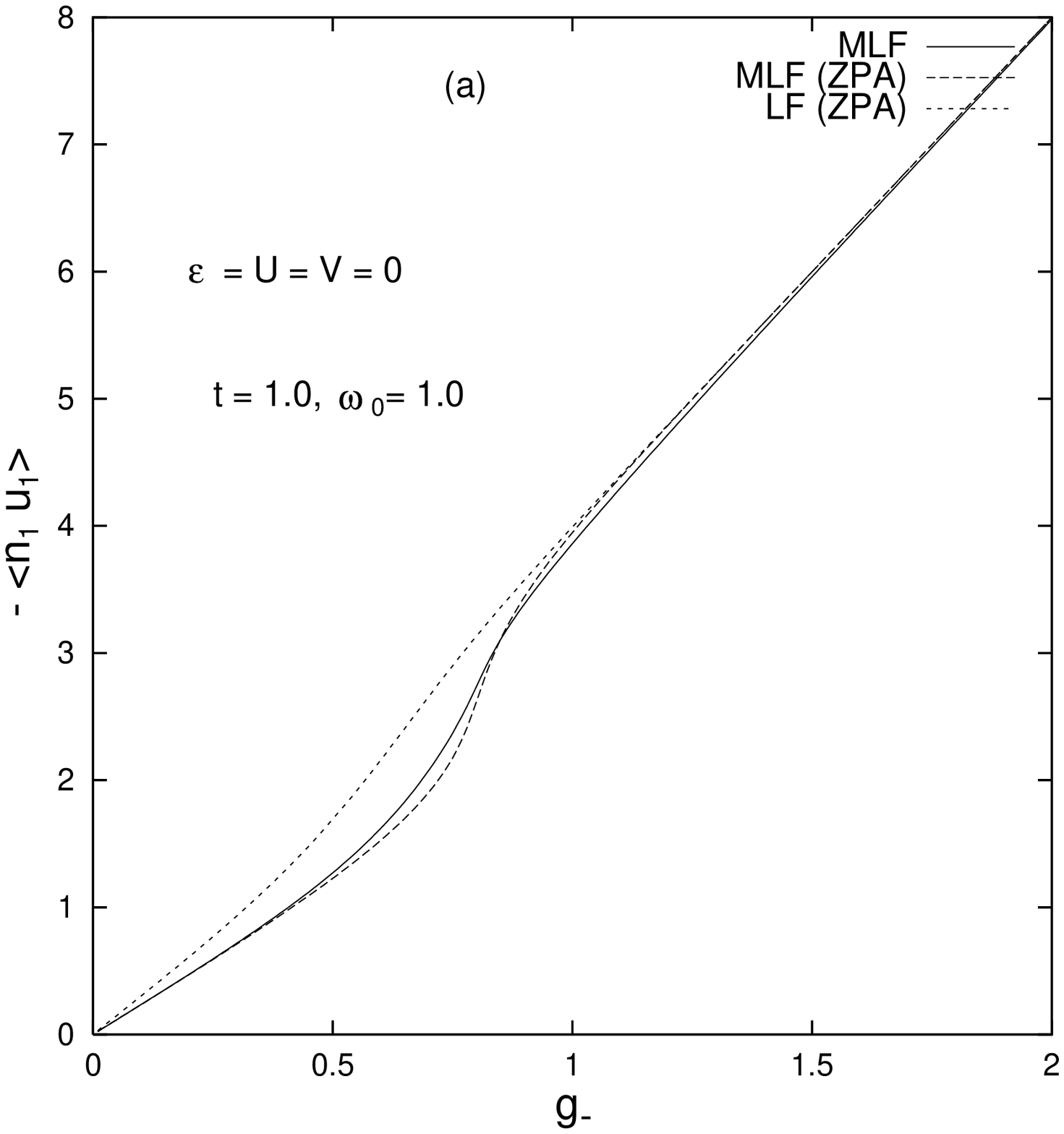,width=3.2in}
\psfig{file=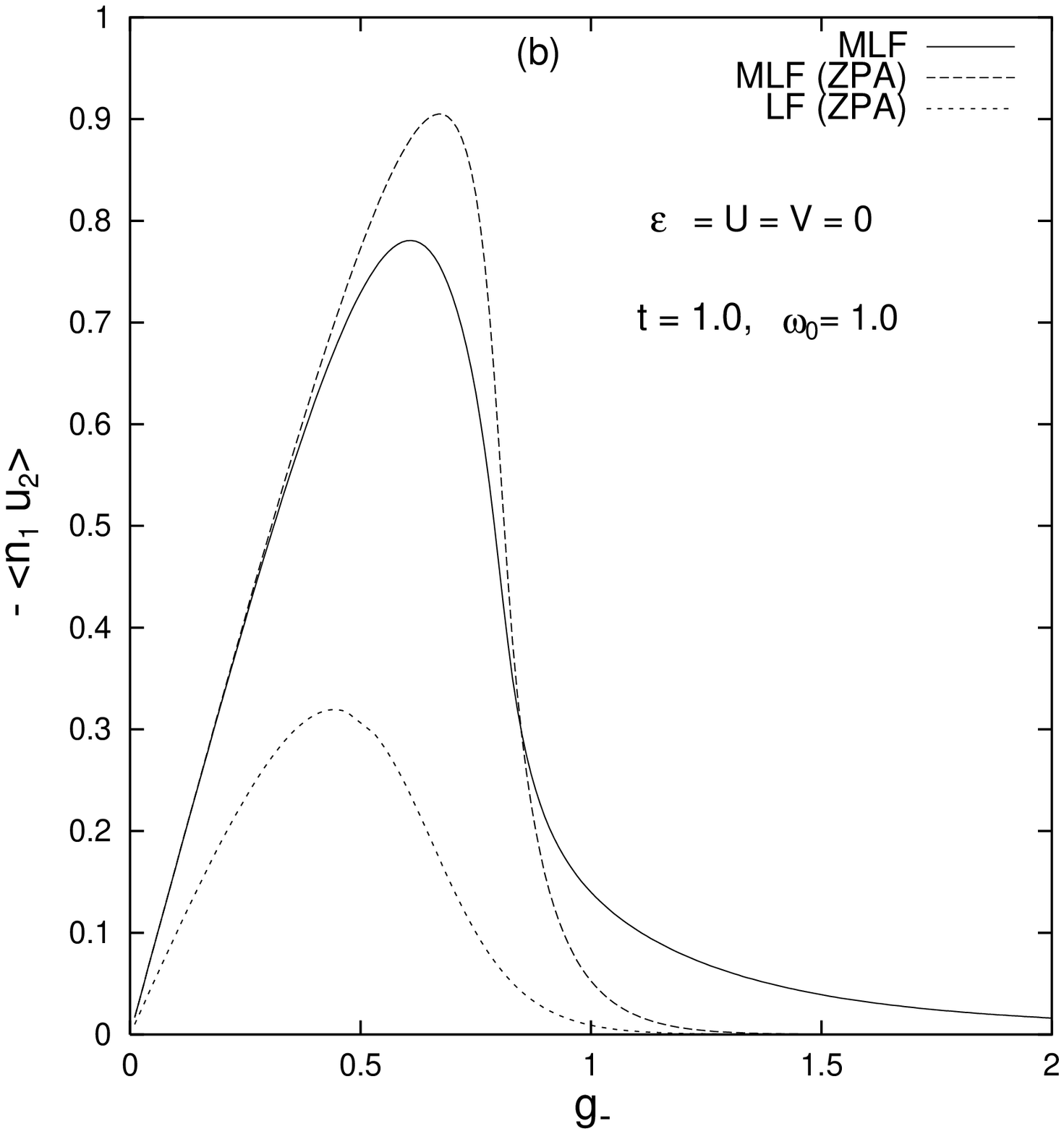,width=3.2in}
\caption{
\label{figure3}
Plot of the correlation functions (a) $\langle n_1u_1 \rangle_{0}$ 
and (b) $\langle n_1u_2 \rangle_{0}$ 
with $g_{-}$ for $t/\omega_0 =1.0$ along with the ZPA results 
within LF and MLF methods.
The solid curve corresponds to that 
obtained within the MLF method considering up to seventh order
correction to the ground-state wave function.
}
\end{figure}
\end{center}
\begin{center}
\begin{figure}
\psfig{file=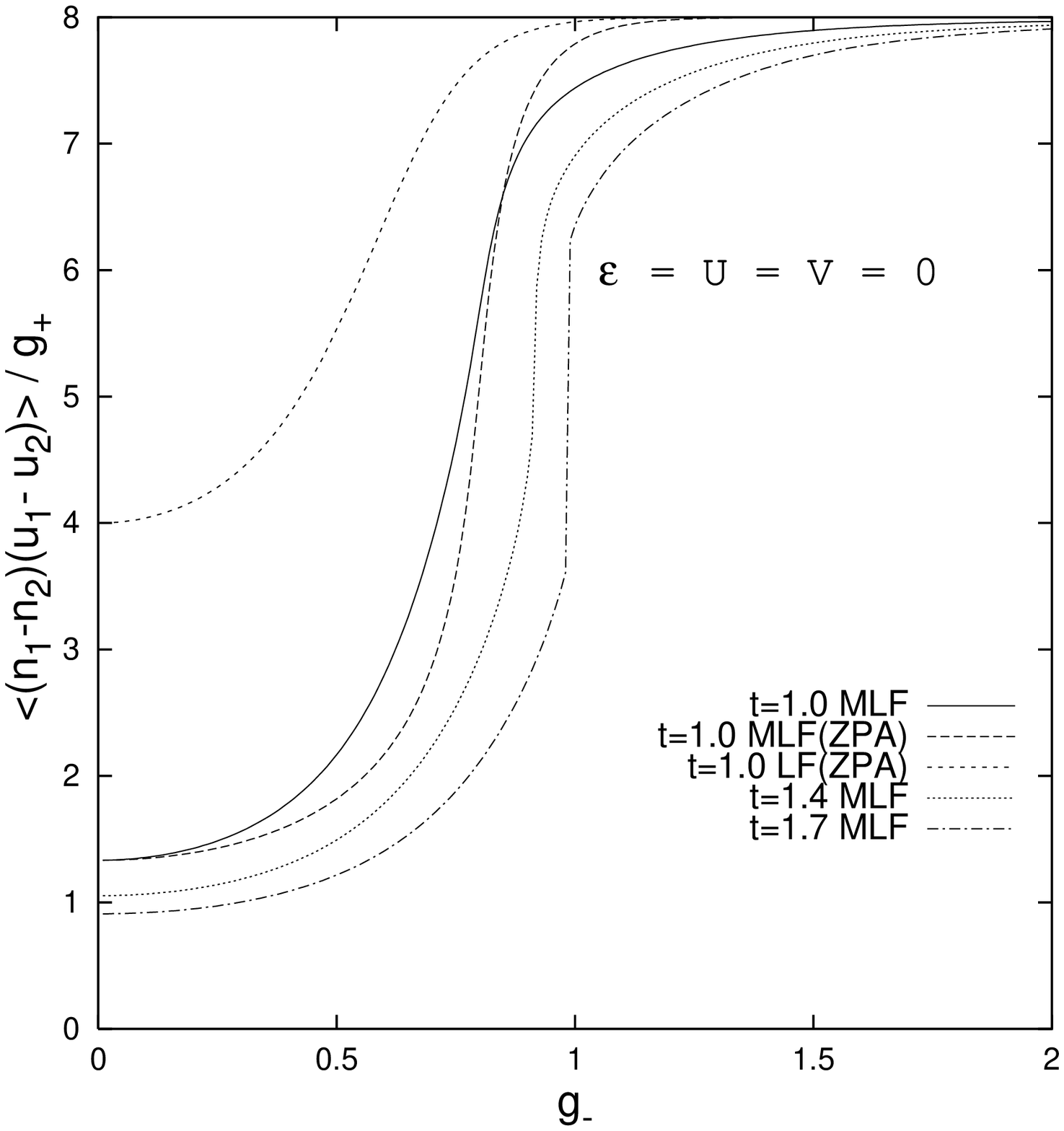,width=3.1in}
\psfig{file=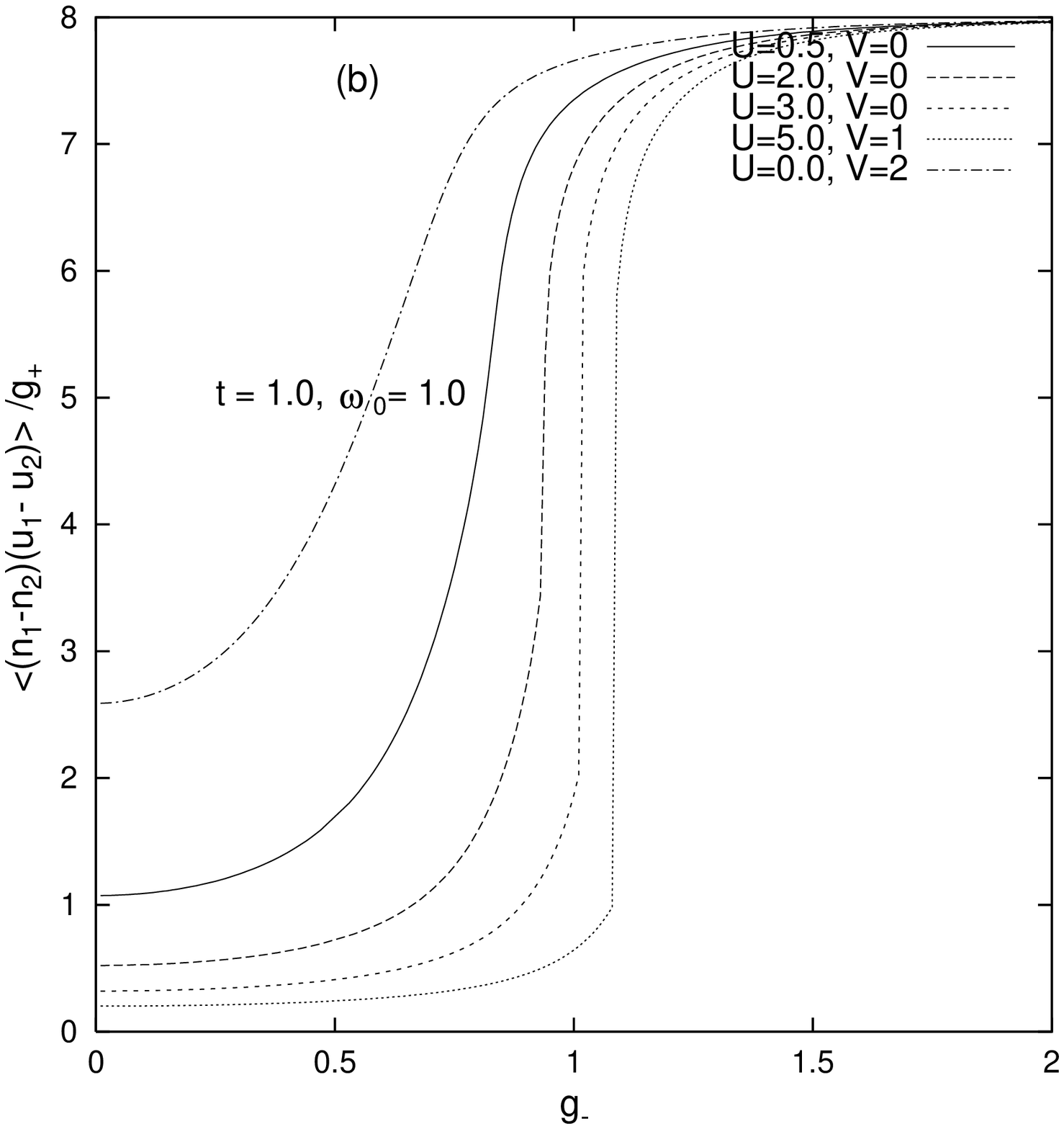,width=3.1in}
\vskip 0.5 cm
\caption{
\label{figure4}
The variations of $ \chi_{12}=  
\langle (n_{1}-n_{2}) (u_{1}-u_{2}) 
\rangle_{0}/g_+$ with $g_-$ within MLF perturbation method
(a) for different values of $t/\omega_0$ and $U=0$; 
the MLF (ZPA) and LF (ZPA) results for
$t/\omega_0=1.0$ are also shown.
(b) for different values of $U$ and $V$ for $t/\omega_0=1.0$.
}
\end{figure}
\end{center}
 To show the crossover from the delocalized (large) to 
localized (small) polaron we plot the quantity 
$\chi_{12}$ (= $\langle (n_{1}-n_{2}) (u_{1}-u_{2}) \rangle_{0}/g_+$)
against $g_-$
in figure 4(a) where the MLF perturbation results
for different values of $t/\omega_0$ are shown.
The ZPA results within MLF and LF methods are also given for
$t/\omega_0=1.0$ to compare them with the exact perturbation results.
The MLF (ZPA) result is very close to the MLF perturbation
result. The MLF perturbation result shows a smooth
crossover from large to small polaron as $e$-ph coupling strength
increases. If we identify the crossover point as the point of inflection 
of the curve obtained within the MLF (perturbation) approach
we find that for $t/\omega_0=1$ the crossover occurs
at a critical $e$-ph coupling $g_-$ ($g_-\sim 0.8$) 
when $U=V=0$. At this point the coefficient of one phonon
state ($C_{N=1}^{1}$ in Eq. (25)) changes its sign.
Similar behaviour has been also observed
in case of our previous single polaron studies \cite{DJC}. The critical
value of the $e$-ph coupling where polaron
crossover occurs increases with increasing value of $t/\omega_0$,
as expected. 

In figure 4(b) we have shown the effect of $e$-$e$ interaction
on the polaron crossover by plotting $\chi_{12}$ against
$g_-$ for several values of $U$ and $V$.
For this two-site two-electron system the polaronic properties
we have studied, depend on $(U-V)$ rather than $U$ and $V$ independently.
From figure 4(b) it is seen that the critical coupling, where the polaron
crossover takes place, increases and the crossover becomes more
abrupt with increasing $(U-V)$. Increasing on-site Coulomb repulsion makes
the crossover region more abrupt and extends the large polaron region.
Similar results were also reported earlier \cite{PF}. The inter-site
Coulomb repulsion, on the other hand, makes the crossover smoother and
the crossover takes place at a lower value of $e$-ph coupling.
\begin{center}
\begin{figure}
\psfig{file=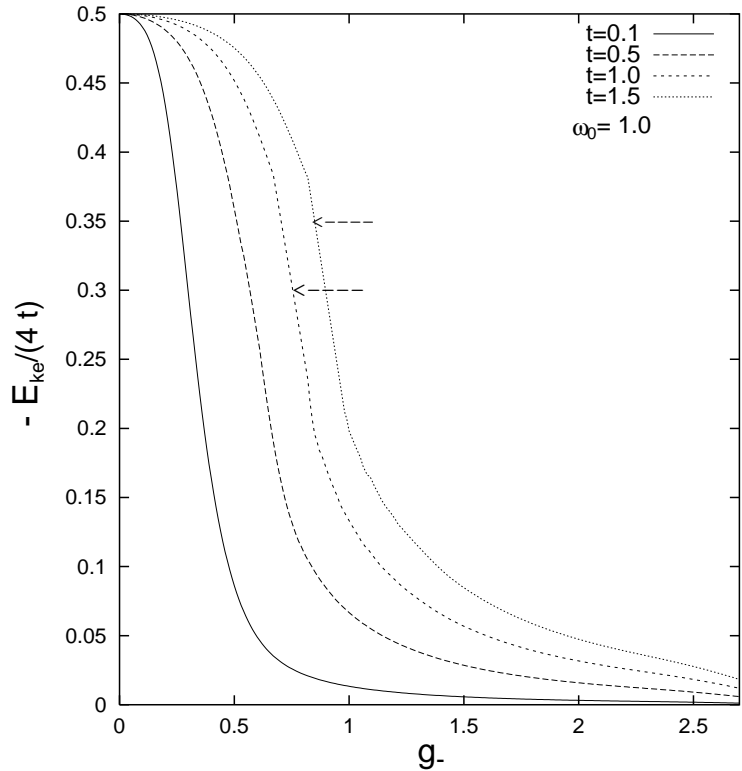,width=4.0in}
\vskip 0.5 cm
\caption{
\label{figure5}
The kinetic energy of the system, 
obtained using the seventh order of perturbation within MLF method,  
as a function of $g_-$ for different values $t/\omega_0$ 
(=0.1, 0.5, 1.0 and 1.5). 
}
\end{figure}
\end{center}
\begin{center}
\begin{figure}
\psfig{file=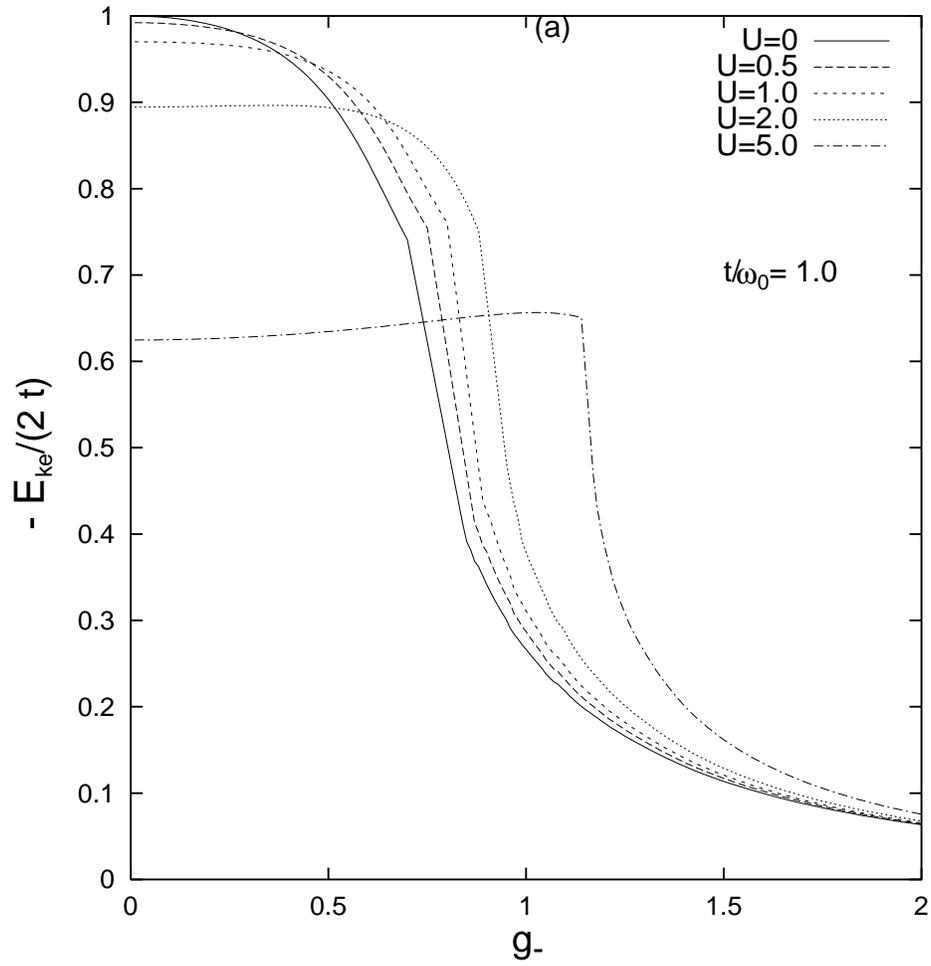,width=5.0in}
\vskip 0.5 cm
\caption{
\label{figure6}
For $t/\omega_0=1.0$, the kinetic energy of the system, obtained by the MLF 
perturbation method, as a function of $g_-$ for different values of
$U$.
}
\end{figure}
\end{center}
 The variation of the kinetic energy of the system with 
$e$-ph coupling strength $g_-$ is shown in figure 5 for $U=~V=~0$ and for 
different values of $t/\omega_0$ (= 0.1, 0.5, 1.0 and 1.5). We find that  
our perturbation results for the kinetic energy match exactly with 
the results obtained by exact diagonalization study \cite{MR} for 
the corresponding set of parameters. It may be mentioned that for 
$t/\omega_0=$ 1.0, 1.5 the convergence is not good in a very small 
region of $g_-$ around the points indicated by the arrows in the figure. 
Except these narrow regions the convergence of our perturbation method 
is excellent in the rest of the parameter space we studied.

 It is well known that the kinetic energy of a tight-binding electronic 
system is suppressed by the introduction of either $e$-ph coupling
or by Coulomb repulsion. An important question one may ask what
is the simultaneous role of the 
Coulomb repulsion and the $e$-ph interaction in lowering
the kinetic energy of the system. To address this question we 
have plotted the kinetic energy against $e$-ph coupling strength for 
different values of $U$ for $t/\omega_0= 1.0$ in figure 6.
It is seen from figure 6 that for large $U$ values there is a flat 
region where the kinetic energy remains unaffected by increase of 
$e$-ph coupling. Higher the value of $U$ larger is the flat region.
Beyond this flat region the kinetic energy is suppressed 
almost exponentially with increasing $g_-$. The occurance of this flat region
and then a rapid fall in the kinetic energy for higher values of $U$
may be understood from the results of figure 4(b) where it is found
that for larger values of $U$ the polaron crossover is abrupt but there is an
extended large polaron region (in $g_-$ space). The value of
$\chi_{12}$, which is a measure of the difference in the lattice distortions
produced at the charge residing site and the neighbouring site, is small
and almost
constant in this large polaron region. As a result a flat behaviour in the
kinetic energy is obtained. Beyond a critical value of $g_-$ an abrupt
increase in $\chi_{12}$ would lead to a sharp fall in the kinetic energy,
which is evident in figure 6. 
\begin{center}
\begin{figure}
\psfig{file=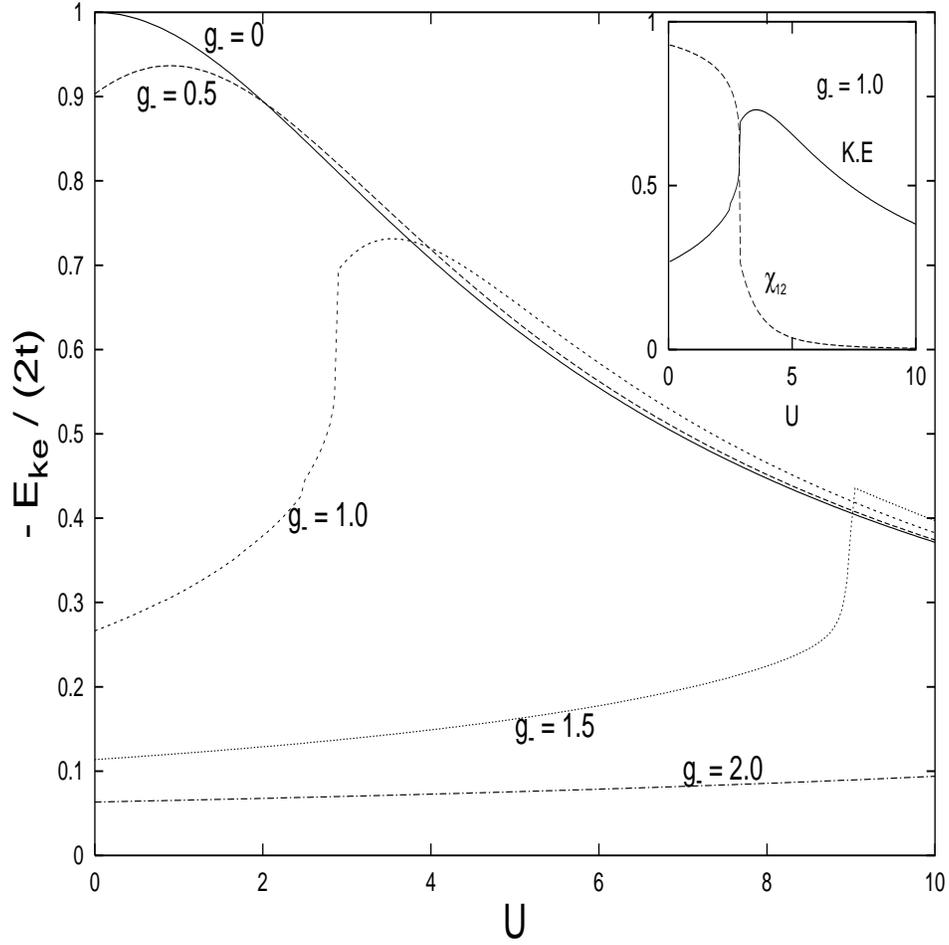,width=5.0in,height=5.0in,angle=-90}
\vskip 0.5 cm
\caption{
\label{figure7}
The kinetic energy of the system, obtained by the MLF 
perturbation method, as a function of $U$ 
for different values of $g_-$ for $t/\omega_0=1.0$; 
The inset figure shows the variation of $\chi_{12}/8$ with $U$
for $g_-=1.0$.
}
\end{figure}
\end{center}
\begin{center}
\begin{figure}
\psfig{file=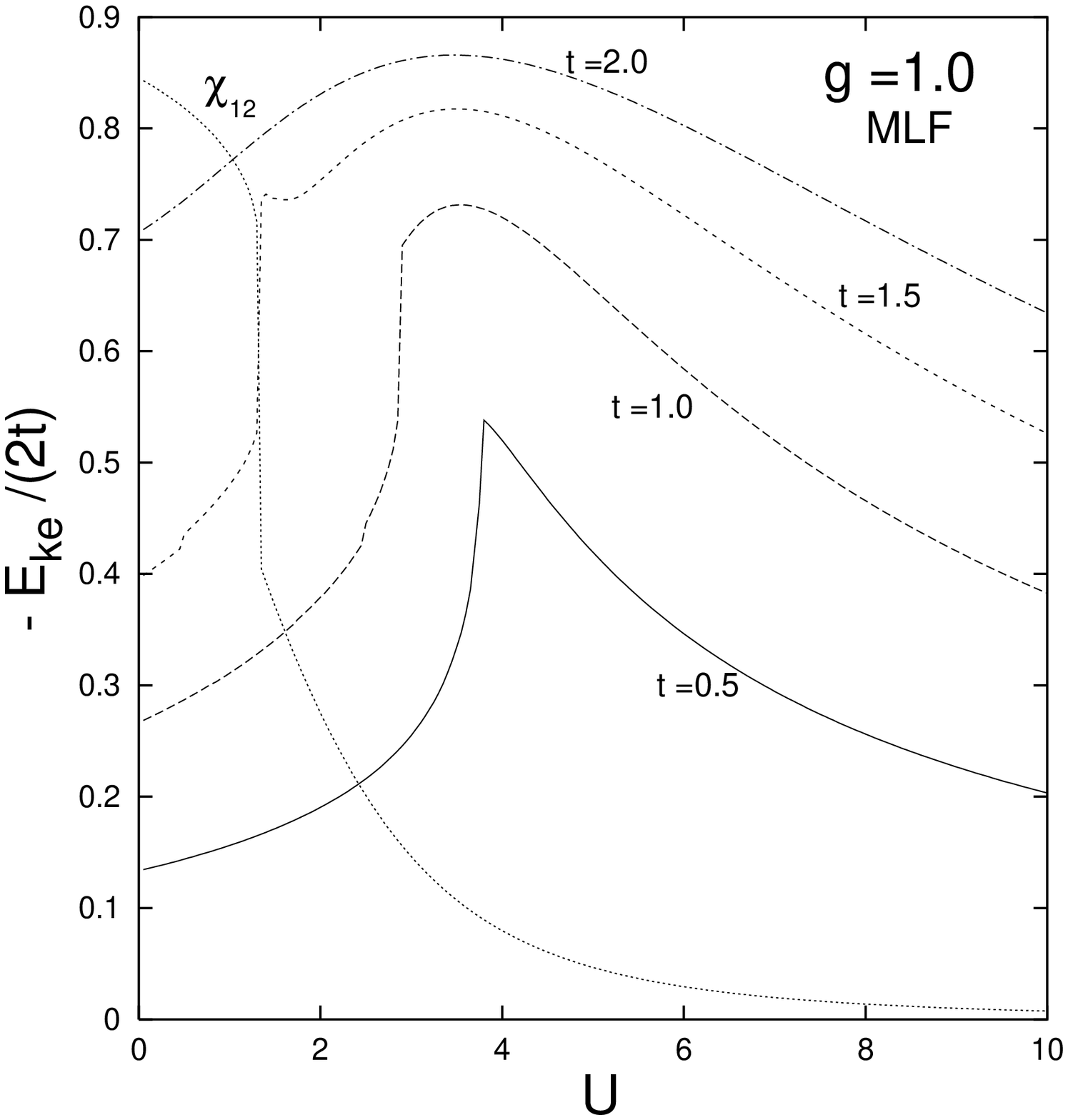,width=5.1in}
\vskip 0.5 cm
\caption{
\label{figure8}
The kinetic energy of the system, obtained by the MLF 
perturbation method, as a function of $U$ for a particular
value of $e$-ph coupling ($g_-=1.0$)
for different values of $t/\omega_0$ (=0.5, 1.0, 1.5 and 2.0).
$\chi_{12}/8$ is also plotted for $t/\omega_0$ =1.5.
}
\end{figure}
\end{center}
 In figure 7 the variation of the kinetic energy 
with $U$ for different values of $g_-$ is shown. It is found that 
for $g_-=1.0$ and $1.5$ there is a region where the kinetic energy 
increases with $U$ in contrast to the usual behaviour that the 
kinetic energy is suppressed by the Coulomb correlation. The 
latter behaviour is of course seen in major parameter space. 
To understand this unusual feature of increasing
kinetic energy with $U$ for finite $g_-$, we plot the variation of
$\chi_{12}$ as well as the kinetic energy with $U$ in the inset of figure
7 for $g_-=1.0$. It is readily seen that as $U$ increases the correlation
function
$\chi_{12}$ decreases which results in an increase in the kinetic energy.
At the crossover region, $\chi_{12}$ shows a sharp fall which is reflected
by the abrupt rise in the kinetic energy. For large values of $U$,
$\chi_{12}$ is small and the polaron is a large one. In this case
the lattice distortions uniformly spread throughout the
whole lattice (hence the charge particle 
behaves more like a normal electron) and
the kinetic energy is suppressed with increasing $U$ as commonly
expected.

 In figure 8 the variation of the kinetic energy 
with $U$ for different values of $t/\omega_0$ is shown
for a fixed value of $e$-ph coupling ($g_-=1.0$). The kinetic energy 
increases with $U$, shows a peak and then decreases. Similar 
behaviour is also observed in figure 7. The peak is broader for
higher values of $t$.
The initial enhancement of the kinetic energy with $U$ is due 
to rapid spread up of the polaron size or decrease in the value of 
$\chi_{12}$ whereas in the large polaron region $U$ plays its 
conventional role of suppressing the kinetic energy. With increasing 
$t$ the kinetic energy increases as well as the large polaron region 
is extended. 
We have also plotted $\chi_{12}$ in the same figure for $t/\omega_0=1.5$
to show the rise in the kinetic energy is related to rapid fall 
in $\chi_{12}$.
For $t/\omega_0=2.0$, the polaron is a large one for $g_-=1$ 
and so the small polaron region is not seen for this case. 

 In summary we apply the MLF perturbation method on a two-site
two-electron Holstein model taking the diagonal part of the Hamiltonian
(in the momentum space of the MLF basis) as the unperturbed 
Hamiltonian. A good convergence is achieved by our approach.
The ground state energy and the kinetic energy obtained 
within this method match exactly with those obtained by exact 
diagonalization study. The behaviour of different correlation functions 
involving charge and the lattice distortions are reported
as a function of $e$-ph coupling. Effect of the Coulomb repulsion and 
the adiabaticity parameter on the large to small polaron crossover 
is studied. On-site Coulomb repulsion makes the polaron crossover 
more abrupt. 
Our study on the kinetic energy in presence of both $e$-ph interaction 
and Coulomb repulsion shows that for large values of $U$, the 
kinetic energy is not suppressed by $e$-ph coupling in a range
which is wider larger the value of $U$. In this  
range the polaron is a large one with almost an uniform value of 
$\chi_{12}$. This makes the kinetic energy insensitive to the 
e-ph coupling. When $e$-ph coupling crosses a critical value a rapid 
crossover to a small polaron occurs and in this region the $e$-ph
coupling is effective in suppressing the kinetic energy.   
On the other hand for finite $e$-ph coupling the 
kinetic energy increases initially with $U$ and then decays. This is
ascribed to the fact that 
the former region corresponds to a small polaron or a crossover region where 
the main role of $U$ is to 
spread the size of the polaron and make it more delocalized. 
This results in an 
increase in the kinetic energy whereas for large $U$ values the polaron 
is a large one and the on-site correlation suppresses the kinetic energy in its
usual way.

\end{document}